# Practical Use of Windows Data Collector: Development Process and Testing Analysis

Daniel Elambo Atonge
*Innopolis University*
Innopolis, Russia
d.atonge@innopolis.ru

Shokhista Ergasheva
*Innopolis University*
Innopolis, Russia
s.ergasheva@innopolis.university

Artem Kruglov
*Innopolis University*
Innopolis, Russia
a.kruglov@innopolis.ru

Giancarlo Succi *Innopolis University* Innopolis, Russia
g.succiv@innopolis.ru

*Abstract*—The paper demonstrates the Windows data collector development process with the built back-end from the require ments gathering stage till the implementation and testing phase. Each phase throughout the development life cycle of the system is defined in details. The whole system idea and the objectives behind developing this kind of framework is described in earlier papers that creates the importance of introducing the background before reading this paper. The detailed information about the data collector features that are provided and their appropriate testing types, documentation are demonstrated thoroughly. Be sides, the process of development includes both design and overall system architecture description.

## I. INTRODUCTION

During the research survey in the field of embedded systems referring the energy related metrics collection it has been an evidence that the battery draining applications result in overall bad user experience and dissatisfied users [1]. The optimal battery usage (energy usage) is an important aspect that every client must consider nowadays [2]–[5]. Application energy consumption is dependent on a wide variety of system resources and conditions. Energy consumption depends on, but is not limited to, the processor the device uses, memory archi tecture, the storage technologies used, the display technology used, the size of the display, the network interface that you are connected to, active sensors, and various conditions like the signal strength used for data transfer, user settings like screen brightness levels, and many more user and system settings [6]. The windows data collector is a tool for estimating the software development process efficiency by collecting process metrics non invasively [7]–[11]. In addition, while estimating the overall process efficiency, the energy consumption metrics are the main novelty of our system.

Modeling the energy consumption of applications, gathering valid data from active and passive application processes (i.e applications in focus and those out of focus) is a crucial activity which can be used to find correlations and trends in various areas of research such as developer's productivity [1], applications with the highest energy consumption profiles and more. To this aim, there have been proposed hardware

This research project is carried out under the support of the Russian Science Foundation Grant N$^{\underline{o}}$ 19-19-00623.

based tools [12] as well as model based and software based techniques to approximate the actual energy profile of appli cations [13]–[15]. However, all these solutions present their own advantages and disadvantages. Hardware based tools are highly precise, but at the same time, their use is bound to the acquisition of costly hardware components. Model based tools require the calibration of parameters needed to correctly create a model on a specific hardware device [6]. Software based approaches are cheaper and easier to use than hardware based tools, but they are believed to be less precise [12], [16]. Moreover, the collected metrics can serve a variety of other purposes, from planning, to monitoring, and then to revising post-mortem software projects [17]–[33]. To this end, techniques from machine learning and computational intelligence are particularly useful [25], [34].

## II. REQUIREMENTS ENGINEERING

Before the analysis of the data collector and designing its overall architecture, we had a series of requirements both functional and nonfunctional that were set so as to govern the design and implementation of the application. These re quirements established the features and functionalities that the client requires from the application. These requirements also specify the constraints under which our system should operate and will be developed. The functional requirements are the following:

• The list of collected metrics should be easily modifiable; • The metrics should be collected and sent to the back-end database automatically after authorization;
• The time interval to send the data to server should be

specifiable;
- The collector should support automatic updates;
- Clients should be able to send error reports;
- Energy-related metrics should be collected;
- Product metrics should be collected;
- Collectors should implement search functionality.

We aim at gathering valuable data related to each running application process. We choose the following quantities to monitor; process name, process id, status (app focus or idle), start time of the process, end time of the process, IP address, mac address, process description, processor, hard disk, mem ory, network and input/output usage.

Data being collected can be subdivided into 2 parts:
- Process metrics which include: process name, process id, status (app focus or idle), start time, end time, IP address, mac address, process description, and
- Energy-related metrics which include: CPU, memory and ram usage.

Similarly, the nonfunctional requirements of the collec tor are modifiability, maintainability, adaptability, security, reusability, and reliability of the system. With these require ments in place, analysis of the developed application and documentation of the design decisions were made.

After the requirements are collected, the analysis phase was took a head on the development life cycle of the system. The iterative analysis process which continues until a preferred and acceptable solution or product emerges was used. During the analysis of our system, factual data was collected (i.e, what is lacking, what was done, what is needed, etc.). Understanding the processes involved, identifying problems and recommend ing feasible suggestions for improving the systems' func tioning was done. This involved studying logical processes, gathering operational data, understanding the information flow, finding out bottlenecks and evolving solutions for overcoming the weaknesses of the system so as to achieve the overall goal for the collector. Furthermore, the subdividing of complex processes involving the entire system and the identification of data store and manual processes were made [35].

### III. DESIGN

Understanding that our application will mostly run in the background as a daemon, we needed simple, scalable, and native approaches to display and collect data.

### A. Use case diagram

During the early stages of the development phase, there was a need for a simple but purposeful representation of the entire system. By means of this representation the followings were specified:
- Specify the context of the system;
- Capture the system requirements;
- Validate the system architecture;
- Drive implementation and generate test cases based on it.

This use case diagram (See Figure 1) describes the best of thinking process where was involved creative skills. The attempt of to give ideas to an efficient system that satisfies the current needs of the clients and the ability to grow in the organizational constraints.

### B. System architecture

The system architecture describes best all the quality at tributes like scalability, fault tolerance and adaptability derived from the requirements. In order to assure that kinds of quality attributes, the system was developed using micro services ar chitecture such as Analytical Service, Administration service, Authorization service and etc.

The data collected is stored in a lightweight database locally on the user's machine for particular period of time and then automatically translated to the server side. This is an efficient way to ensure persistence and integrity in case of unexpected run time error or system shutdown. The local database maintains collected data and checks sent data on each successful transfer to the back-end to avoid data duplication and inconsistencies.

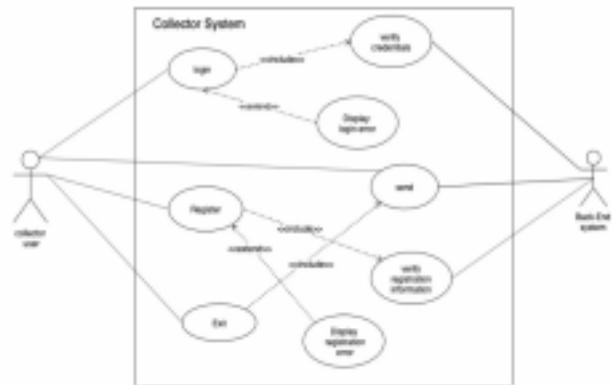

Fig. 1. Use case diagram of Windows Collector

For an initial deployment of the solution, there is a server that will host all the services shown in the Figure 2.

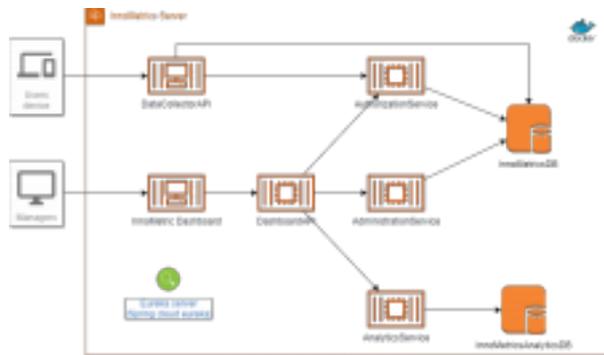

Fig. 2. System architecture

## IV. Implementation

To provide the system help the developers and to start the data collection process and validate the data model, the system with a monolithic API was developed. Besides, in order to provide the system with the required nonfunctional requirements, the second version of the system with different components, where each service has a particular purpose. The system was created based on microservices with two main access points and a series of specialized services such as:

- Analytic Service - to be able to extract information from data collected and to carry out the data transformation referring the model that is suitable for end user.
- Administration services - to provide functionalities that allow to make configurations in the system as a user and administration.
- Authorization service - to authenticate and authorize the users, to generate and validate tokens to process requests from various services.

The architecture of the system included decision making process, to achieve the stated requirements. The program, in cluding submodules and services, were written as to coordinate data collection, data storage, data transfer and control the entire processes built into our system. The code was well written to reduce the testing and maintenance effort [35]. This helps in fast development, maintenance, and future changes if required. Visual Studio Community 2019 version and the C# programming language was used [36], [37].

## V. Testing

After developing the various features of the windows collec tor and sticking to the constraints specified in the requirements, it is necessary to carry out some testing with the back-end system developed to verify whether our system actually works without crashing, whether it sends the collected data efficiently without loss and whether it satisfies all the constraints stated in the requirements document. From the windows data col lector perspective, the back-end has to carry out several tasks including but not limited to the following:

• Account verification (at login and
registration) • Receive data sent by a user

Before deciding to install our system and put it into opera tion, a test run of the system was done removing all the bugs as necessary. The output of all our tests matched the expected results. We ran the two categories of test:

- Program test. Referring to our requirements for the expected results of our system, coding, compiling and running our executable was the routine. After our locally based col lector application was up and running, each use case of our system was tested to match the expected outcome. We car ried out various verification and noted unforeseen happenings which were eventually corrected.

- System test, which implies running of the overall system with the back-end and making sure it meets the specified requirements were done. Our system was fully developed and ready for usage by clients.

After testing our data collector for a long period of time, and running queries in the back-end to export collected data into CSV format, we obtain the following result (Figure 3). Added to the previous system test, this validates that our developed collector successfully communicates with the back end without loss of data.

## VI. Conclusion

Battery draining applications result in bad user experience and dissatisfied users [1]. Optimal battery usage (energy usage) is an important aspect that every client must consider nowadays [38]. Application energy consumption is dependent on a wide variety of system resources and conditions. Energy consumption depends on, but is not limited to, the processor the device uses, memory architecture, the storage technologies used, the display technology used, the size of the display, the network interface that you are connected to, active sensors, and various conditions like the signal strength used for data transfer, user settings like screen brightness levels, and many more user and system settings [6].

Fig. 3. Back-end data

For precise energy consumption measurements, one needs specialized hardware. While they provide the best method to accurately measure energy consumption on a particular device, such a methodology is not scalable in practice, especially if such measurements have to be made on multiple devices. Even then, the measurements by themselves will not provide much insight into how your application contributed to the battery drain, making it hard to focus on any application optimization efforts.

The collector aims at enabling users to estimate their appli cation's energy consumption without the need for specialized hardware. Such estimation is made possible using a software power model that has been trained on a reference device representative of the low powered devices.

Using the PerformanceCounter, PerformanceCounterCate gory and many more related classes (made available by the .NET Framework) [36], [37], the energy usage can be computed. Performance counter(s) provided information such as CPU time, Total Processor Time per process, CPU usage, Memory usage, network usage and more. The MSDN docu mentation was used to better understand how we could utilize the available components in attaining our goal (collecting energy consumption metrics).

Need to mention that it is difficult to match up constantly changing application process IDs and names. Also, imperfec tions in the designed power model as energy consumption depends on a variety of factors not limited to those we can collect using these available performance classes.

In this report, we have presented the requirements that play the crucial role in the system development, the design and the overall system architecture details, a deeper view of how the data collector operates with the back-end, analyzed the use cases, the testing phase with some results displayed of testing, and stated the drawback experienced during this process.

The plan is to release the collector with some open source license to ensure its maximal distribution [39]–[43].

ACKNOWLEDGMENTS

We thank AK Bars Digital Technologies for their help in this work. This study was supported by Russian Science
Foundation Grant №19-19-00623.

REFERENCES

[1] Ah-Lian Kor, Colin Pattinson, Ismail Imam, Ibrahim AlSaleemi, and Oluwafemi Omotosho. Applications, energy consumption, and mea surement. In *2015 International Conference on Information and Digital Technologies*. IEEE, July 2015.

[2] Luis Corral, Alberto Sillitti, Giancarlo Succi, Alessandro Garibbo, and Paolo Ramella. Evolution of Mobile Software Development from Platform-Specific to Web-Based Multiplatform Paradigm. In *Proceed ings of the 10th SIGPLAN Symposium on New Ideas, New Paradigms, and Reflections on Programming and Software*, Onward! 2011, pages 181–183, New York, NY, USA, 2011. ACM.

[3] Luis Corral, Anton B Georgiev, Alberto Sillitti, and Giancarlo Succi. A method for characterizing energy consumption in Android smartphones. In *Green and Sustainable Software (GREENS 2013), 2nd International Workshop on*, pages 38–45. IEEE, May 2013.

[4] Luis Corral, Anton B. Georgiev, Alberto Sillitti, and Giancarlo Succi. Can execution time describe accurately the energy consumption of mobile apps? An experiment in Android. In *Proceedings of the 3rd International Workshop on Green and Sustainable Software*, pages 31– 37. ACM, 2014.

[5] Luis Corral, Alberto Sillitti, and Giancarlo Succi. Software Assurance Practices for Mobile Applications. *Computing*, 97(10):1001–1022, October 2015.

[6] Erik Jagroep, Giuseppe Procaccianti, Jan Martijn van der Werf, Sjaak Brinkkemper, Leen Blom, and Rob van Vliet. Energy efficiency on the product roadmap: An empirical study across releases of a software product. *Journal of Software: Evolution and Process*, 29(2):e1852, February 2017.

[7] Tullio Vernazza, Giampiero Granatella, Giancarlo Succi, Luigi Benedi centi, and Martin Mintchev. Defining Metrics for Software Components. In *Proceedings of the World Multiconference on Systemics, Cybernetics and Informatics*, volume XI, pages 16–23, July 2000.

[8] Marco Scotto, Alberto Sillitti, Giancarlo Succi, and Tullio Vernazza. A Relational Approach to Software Metrics. In *Proceedings of the 2004 ACM Symposium on Applied Computing*, SAC '04, pages 1536–1540. ACM, 2004.

[9] Alberto Sillitti, Andrea Janes, Giancarlo Succi, and Tullio Vernazza. Measures for mobile users: an architecture. *Journal of Systems Archi tecture*, 50(7):393–405, 2004.

[10] Marco Scotto, Alberto Sillitti, Giancarlo Succi, and Tullio Vernazza. A non-invasive approach to product metrics collection. *Journal of Systems Architecture*, 52(11):668–675, 2006.

[11] Andrea Janes and Giancarlo Succi. *Lean Software Development in Action*. Springer, Heidelberg, Germany, 2014.

[12] Erik A. Jagroep, Jan Martijn E. M. van der Werf, Ruvar Spauwen, Leen Blom, Rob van Vliet, and Sjaak Brinkkemper. An energy consumption perspective on software architecture. In *Software Architecture*, pages 239–247. Springer International Publishing, 2015.

[13] Alberto Sillitti, Tullio Vernazza, and Giancarlo Succi. Service Oriented Programming: A New Paradigm of Software Reuse. In *Proceedings of the 7th International Conference on Software Reuse*, pages 269–280. Springer Berlin Heidelberg, April 2002.

[14] Justin Clark, Chris Clarke, Stefano De Panfilis, Giampiero Granatella, Paolo Predonzani, Alberto Sillitti, Giancarlo Succi, and Tullio Vernazza. Selecting components in large cots repositories. *Journal of Systems and Software*, 73(2):323–331, 2004.

[15] Andrea Janes, Marco Scotto, Alberto Sillitti, and Giancarlo Succi. A perspective on non invasive software management. In *2006 IEEE Instrumentation and Measurement Technology Conference Proceedings*. IEEE, April 2006.

[16] Hayri Acar, Gulfem Isiklar Alptekin, Jean-Patrick Gelas, and Parisa Ghodous. The impact of source code in software on power consumption. *IJEBM*, 14, 2016.

[17] Giuseppe Marino and Giancarlo Succi. Data Structures for Parallel Execution of Functional Languages. In *Proceedings of the Parallel Architectures and Languages Europe, Volume II: Parallel Languages*, PARLE '89, pages 346–356. Springer-Verlag, June 1989.

[18] Andrea Valerio, Giancarlo Succi, and Massimo Fenaroli. Domain analysis and framework-based software development. *SIGAPP Appl. Comput. Rev.*, 5(2):4–15, September 1997.

[19] Frank Maurer, Giancarlo Succi, Harald Holz, Boris Kotting, Sigrid Goldmann, and Barbara Dellen. Software Process Support over

[19] ... the Internet. In *Proceedings of the 21st International Conference on Software Engineering*, ICSE '99, pages 642–645. ACM, May 1999.

[20] Giancarlo Succi, Luigi Benedicenti, and Tullio Vernazza. Analysis of the effects of software reuse on customer satisfaction in an RPG environment. *IEEE Transactions on Software Engineering*, 27(5):473–479, 2001.

[21] Giancarlo Succi, James Paulson, and Armin Eberlein. Preliminary results from an empirical study on the growth of open source and commercial software products. In *EDSER-3 Workshop*, pages 14–15, 2001.

[22] Giancarlo Succi, Witold Pedrycz, Michele Marchesi, and Laurie Williams. Preliminary analysis of the effects of pair programming on job satisfaction. In *Proceedings of the 3rd International Conference on Extreme Programming (XP)*, pages 212–215, May 2002.

[23] Petr Musílek, Witold Pedrycz, Nan Sun, and Giancarlo Succi. On the Sensitivity of COCOMO II Software Cost Estimation Model. In *Proceedings of the 8th International Symposium on Software Metrics*, METRICS '02, pages 13–20. IEEE Computer Society, June 2002.

[24] James W Paulson, Giancarlo Succi, and Armin Eberlein. An empirical study of open-source and closed-source software products. *IEEE Transactions on Software Engineering*, 30(4):246–256, 2004.

[25] Witold Pedrycz and Giancarlo Succi. Genetic granular classifiers in modeling software quality. *Journal of Systems and Software*, 76(3):277–285, 2005.

[26] Marco Ronchetti, Giancarlo Succi, Witold Pedrycz, and Barbara Russo. Early estimation of software size in object-oriented environments a case study in a cmm level 3 software firm. *Information Sciences*, 176(5):475–489, 2006.

[27] Raimund Moser, Witold Pedrycz, and Giancarlo Succi. A Comparative Analysis of the Efficiency of Change Metrics and Static Code Attributes for Defect Prediction. In *Proceedings of the 30th International Conference on Software Engineering*, ICSE 2008, pages 181–190. ACM, 2008.

[28] Raimund Moser, Witold Pedrycz, and Giancarlo Succi. Analysis of the reliability of a subset of change metrics for defect prediction. In *Proceedings of the Second ACM-IEEE International Symposium on Empirical Software Engineering and Measurement*, ESEM '08, pages 309–311. ACM, 2008.

[29] Bruno Rossi, Barbara Russo, and Giancarlo Succi. Modelling Failures Occurrences of Open Source Software with Reliability Growth. In *Open Source Software: New Horizons - Proceedings of the 6th International IFIP WG 2.13 Conference on Open Source Systems, OSS 2010*, pages 268–280, Notre Dame, IN, USA, May 2010. Springer, Heidelberg.

[30] Witold Pedrycz, Barbara Russo, and Giancarlo Succi. A model of job satisfaction for collaborative development processes. *Journal of Systems and Software*, 84(5):739–752, 2011.

[31] Alberto Sillitti, Giancarlo Succi, and Jelena Vlasenko. Understanding the Impact of Pair Programming on Developers Attention: A Case Study on a Large Industrial Experimentation. In *Proceedings of the 34th International Conference on Software Engineering*, ICSE '12, pages 1094–1101, Piscataway, NJ, USA, June 2012. IEEE Press.

[32] Enrico Di Bella, Alberto Sillitti, and Giancarlo Succi. A multivariate classification of open source developers. *Information Sciences*, 221:72–83, 2013.

[33] Irina D Coman, Pierre N Robillard, Alberto Sillitti, and Giancarlo Succi. Cooperation, collaboration and pair-programming: Field studies on backup behavior. *Journal of Systems and Software*, 91:124–134, 2014.

[34] Witold Pedrycz, Barbara Russo, and Giancarlo Succi. Knowledge Transfer in System Modeling and Its Realization Through an Optimal Allocation of Information Granularity. *Appl. Soft Comput.*, 12(8):1985–1995, August 2012.

[35] John Dooley. *Software Development and Professional Practice*. Apress, 2011.

[36] Performance counters. https://docs.microsoft.com/en us/windows/win32/perfctrs/performance-counters-portal, December 2019.

[37] Performance counters in the .NET framework. https://docs.microsoft.com/en-us/dotnet/framework/debug-trace profile/performance-counters, December 2019.

[38] S Ergasheva, I Khomyakov, A Kruglov, and G Succil. Metrics of energy consumption in software systems: a systematic literature review. *IOP Conference Series: Earth and Environmental Science*, 431:012051, February 2020.

[39] Jeremy Kivi, Darlene Haydon, Jason Hayes, Ryan Schneider, and Gian carlo Succi. Extreme programming: a university team design experience. In *2000 Canadian Conference on Electrical and Computer Engineering. Conference Proceedings. Navigating to a New Era (Cat. No.00TH8492)*, volume 2, pages 816–820 vol.2, May 2000.

[40] Gyorgy L Kovács, Sylvester Drozdik, Paolo Zuliani, and Giancarlo ´ Succi. Open Source Software for the Public Administration. In *Proceedings of the 6th International Workshop on Computer Science and Information Technologies*, October 2004.

[41] Etiel Petrinja, Alberto Sillitti, and Giancarlo Succi. Comparing Open BRR, QSOS, and OMM assessment models. In *Open Source Software: New Horizons - Proceedings of the 6th International IFIP WG 2.13 Conference on Open Source Systems, OSS 2010*, pages 224–238, Notre Dame, IN, USA, May 2010. Springer, Heidelberg.

[42] Brian Fitzgerald, Jay P Kesan, Barbara Russo, Maha Shaikh, and Giancarlo Succi. *Adopting open source software: A practical guide*. The MIT Press, Cambridge, MA, 2011.

[43] Bruno Rossi, Barbara Russo, and Giancarlo Succi. Adoption of free/libre open source software in public organizations: factors of impact. *Information Technology & People*, 25(2):156–187, 2012.